%% file: asteroFLAG_arxiv_v1.tex
\documentclass[a4paper,7pt,twoside]{CoAst}
\usepackage{epsf,graphicx,natbib,fancyhdr}

\input{CoAst_regularlayo}

\input{journals}
\usepackage[margin=3cm,includehead,includefoot,heightrounded,a4paper]{geometry}

\usepackage{amssymb,amsmath} 

\bibpunct[; ]{(}{)}{;}{a}{}{}

\def\sun{\odot}

\begin{document}
\sf

\chapterCoAst{Automated extraction of oscillation parameters for Kepler observations of solar-type stars}
%paper titel and page heading for even pages
{D.\,Huber, D.\,Stello, T.R.\,Bedding, W.J.\,Chaplin et al.} %page heading for odd pages
{Automated extraction of oscillation parameters for Kepler observations of solar-type stars}
\Authors{Daniel Huber$^{1}$, Dennis Stello$^1$, Timothy R. Bedding$^1$, William J. Chaplin$^{2}$, 
Torben Arentoft$^{3}$, Pierre-Olivier Quirion$^{3}$ and Hans Kjeldsen$^{3}$} 
\Address{$^1$ Sydney Institute for Astronomy (SIfA), School of Physics, University of Sydney, NSW 2006, Australia\\
$^2$ School of Physics and Astronomy, University of Birmingham, Edgbaston, Birmingham, B15 2TT, UK \\
$^3$ Department of Physics and Astronomy, University of Aarhus, DK-8000 Aarhus C, Denmark
}

\noindent
\begin{abstract}
The recent launch of the Kepler space telescope brings the opportunity to
study oscillations systematically in large numbers of solar-like stars. In the framework of the asteroFLAG 
project, we have developed an automated pipeline to estimate global oscillation parameters, 
such as the frequency of maximum power ($\nu_{\rm max}$) and the large 
frequency spacing ($\Delta\nu$), for a large number of time series. We present an 
effective method based on the autocorrelation function to find excess power and use a scaling 
relation to estimate granulation timescales as initial conditions for background modelling. We derive 
reliable uncertainties for $\nu_{\rm max}$ and $\Delta\nu$ through extensive simulations. We have tested 
the pipeline on about 
2000 simulated Kepler stars with magnitudes of $V$\,$\sim$\,7--12 and were able to correctly determine 
$\nu_{\rm max}$ and $\Delta\nu$ for about half of the sample. For about 20\%, the returned large frequency 
spacing is accurate 
enough to determine stellar radii to a 1\% precision. We conclude that the methods presented here 
are a promising approach to process the large amount of data expected from Kepler.

\end{abstract}

\section*{1 Introduction}
Stellar oscillations are a powerful tool to study the interiors of stars and 
to determine their fundamental parameters. Until recently, the detection of oscillations in 
solar-type stars has been possible only for a handful of bright stars \citep[see, e.g.,][]{bedding2008}. 
With the launch of the space telescopes CoRoT 
\citep{baglin} and Kepler \citep{borucki}, however, this situation is changing. 
In pursuing its main mission goal of detecting transits of extrasolar planets around solar-like stars, 
Kepler will photometrically monitor thousands of stars for a period of up to four years. 
Asteroseismology will allow us to determine radii of exoplanet host stars 
\citep[Kjeldsen et al. 2009]{daals,stello2007}, and also to study 
oscillations systematically in a large number of solar-type stars for the first time.

To deal with the amount of data that Kepler is expected to return, automatic analysis pipelines 
are needed. Such algorithms have already been successfully applied to CoRoT exofield 
data to study oscillations in red giants \citep{hekker2009}. For Kepler, the development of analysis 
tools has been carried out in the framework of the asteroFLAG project \citep{chaplin2008_2, mathur} through 
so-called Hare \& Hounds exercises, in which one group (the Hounds) analyse simulated data produced 
by others (the Hares) without knowing the parameters on which the simulations are based. 
\citet{chaplin2008} presented the results of the first exercise, which concentrated on a few stars 
simulated at different evolutionary stages with various apparent magnitudes and a time base of 4 years 
(as expected for a full-length Kepler time series). The results were then used in a second exercise 
to test the ability to determine radii using stellar models \citep{stello_radius}.

In this paper, we describe an automated pipeline to extract oscillation parameters such as the 
frequency of maximum power ($\nu_{\rm max}$) and the mean large frequency spacing ($\Delta\nu$). 
We apply it to a large sample of simulated time series that are based on stellar parameters of real 
stars selected for the Kepler asteroseismology survey phase. During this phase, which will occupy the first nine 
months of Kepler science operations, about 2000 stars will be monitored for one month each. These data are 
intended to characterise a large number of solar-like stars and the results will be used to verify 
the Kepler Input Catalog \citep{brown_KIC}, as well as to select high-priority targets to be 
observed for the entire length of the mission.

\section*{2 asteroFLAG simulations}
The simulated Kepler light curves were produced using a combination of the asteroFLAG 
simulator (Chaplin et al., in preparation) and the KASOC simulator (T. Arentoft, unpublished). 
All simulations include stellar granulation, activity cycles and 
instrumental noise, as well as oscillation frequencies computed using the stellar evolution and 
pulsation codes ASTEC \citep{astec} and ADIPLS \citep{adipls}, together 
with rotational splitting and theoretical damping rates. To simulate Kepler survey targets, 
fundamental parameters were taken from the Kepler Input Catalog. Next, a model 
within estimated uncertainties of these parameters was chosen for each star. Every simulated 
light curve had a length of one month, with a sampling time of 60 seconds (representative for 
real Kepler time series). In total, 1936 stars in the magnitude range $V$\,$\sim$\,7--12 were simulated, 
and we used this sample to test the pipeline that is described in the next section.

\section*{3 Data analysis pipeline}
The pipeline covers the first basic analysis steps that will be 
performed on the Kepler light curves. These are: (a) estimating the position of power excess 
in the power spectrum, (b) fitting to and correcting for the background, and (c) estimating the mean 
large frequency spacing. Locating the
power excess due to oscillations not only constrains 
fundamental parameters of a star (in particular, its luminosity), but is also crucial for a successful 
automation of subsequent analysis steps. A problem when analysing the oscillation signal is the
non-white background noise due to variability caused by granulation and stellar activity. 
For the analysis of red giants 
observed with CoRoT, \citet{kallinger_RG} used simultaneous fitting of the background 
and the oscillation power excess, with the latter modelled with a Gaussian function. Here, we 
separate these two steps by first locating the power excess region, and then excluding the 
identified region when modelling the background. Finally, the background-corrected spectrum is used 
to estimate the large frequency spacing in the region where the power was located. In the 
following subsections, each of these three analysis steps will be described in detail.

\subsection*{3.1 Locating the power excess}
\label{sec31}
To locate the power excess, we follow a three-step procedure that is demonstrated in Figure \ref{fig01} using a 
30-day VIRGO time series of the Sun \citep{soho}:

\begin{itemize}
\item[(1)] The background is crudely estimated by binning the power spectrum in equal 
logarithmic bins and smoothing the result with a median filter. The optimal width of the bins depends 
on the frequency resolution of the data, and typical values for the 30-day asteroFLAG stars were 
logarithmic bins with a width of 0.005\,log($\mu$Hz). 
\item[(2)] The residual power spectrum, after subtracting this background (Figure \ref{fig01}, top panel), 
is divided into subsets roughly equal to 4$\Delta\nu$ and overlapping by 50\,$\mu$Hz. The mean of 
each subset is subtracted and the absolute 
autocorrelation function (ACF) for each is calculated for a pre-defined range of frequency spacings 
(Figure \ref{fig01}, middle panel). Note that in order to conserve information about the actual power 
level in the power spectrum of a subset, the ACF is not normalised to unity at zero spacing.
\item[(3)] For each subset, represented by its central frequency, we collapse the ACF 
over all frequency spacings (Figure \ref{fig01}, bottom panel).  
We finally fit a Gaussian function to the peak of the collapsed ACF to localise the 
power excess region (thick grey line). We take the centre of the Gaussian 
to be our measurement of the frequency of maximum power, $\nu_{\rm max}$.
\end{itemize}

\figureDSSN{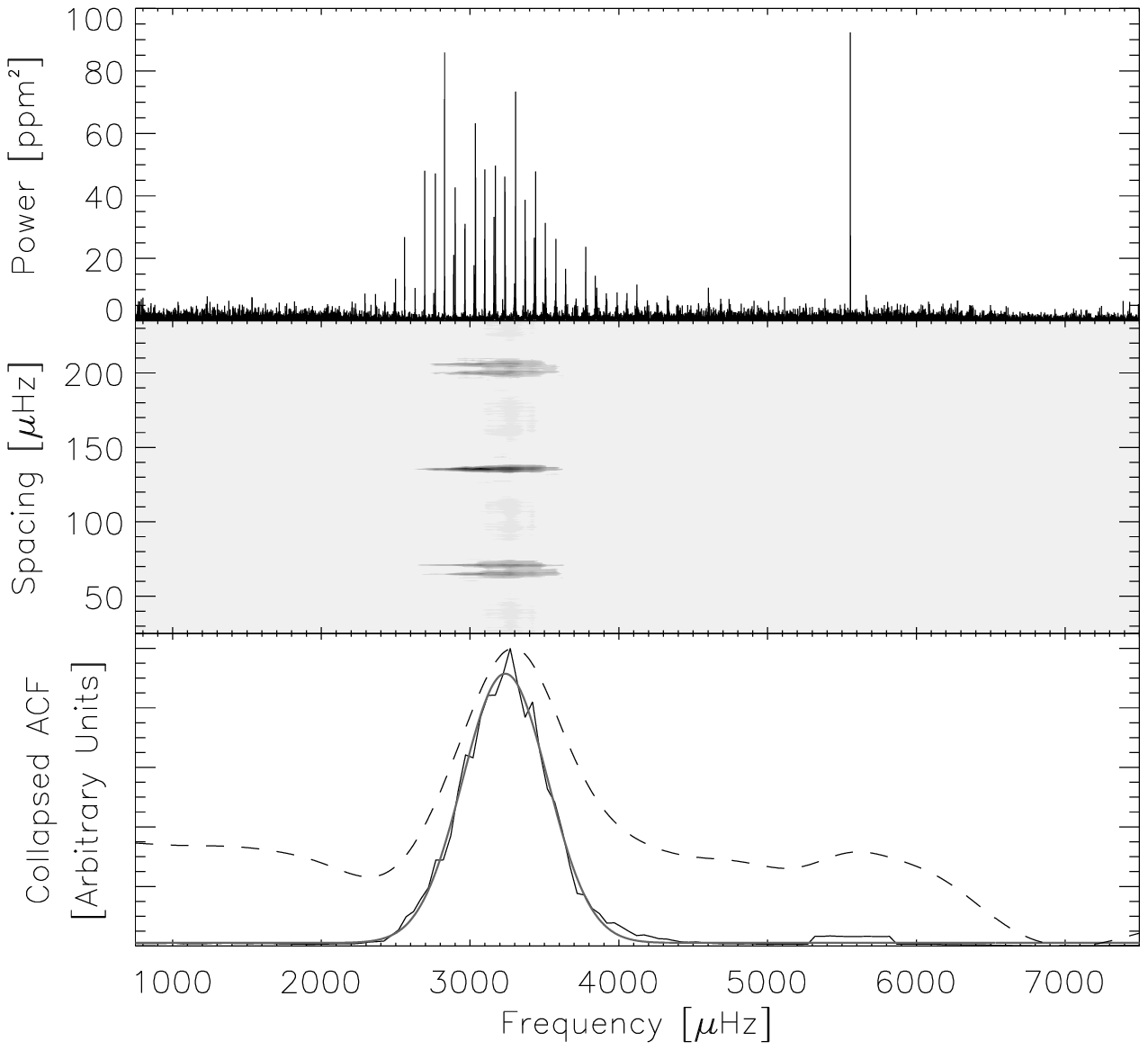}{Procedure for locating the power excess using a 30 day subset of VIRGO photometry. 
Top panel: Background corrected power spectrum. 
Middle Panel:  Autocorrelation as a function of frequency spacing and central frequency of the 
subset at which the correlation is evaluated. Dark colors are regions of high correlation. 
Bottom Panel: Collapsed ACF (black solid line) and smoothed power spectrum (dashed line). The grey 
solid line shows a Gaussian fit 
to the collapsed ACF.}{fig01}{t!}{clip,angle=0,width=100mm}

More precisely, a vertical cut through the middle panel of Figure \ref{fig01} at a given 
frequency is the ACF of the power spectrum subset centered at that frequency. 
In this example, the subset length chosen was 4$\Delta\nu$ ($\sim 540\,\mu$Hz). In applications 
where no estimate for $\Delta\nu$ is available, a range of up to three subset widths are 
applied and the one returning the highest S/N in the collapsed ACF is taken for the $\nu_{\rm max}$ 
estimate. As expected for this example, 
the ACF shows large values at multiples 
of half the large frequency spacing of the Sun ($\sim$\,68\,$\mu$Hz), concentrated at frequencies around 
3\,mHz in the power spectrum. The collapsed ACF in the bottom panel is calculated by 
summing the middle panel vertically. 
For comparison, the dashed line shows the power spectrum smoothed with a Gaussian function with a 
FWHM of 4$\Delta\nu$.

An advantage of this technique over smoothing the power spectrum is that the collapsed ACF is 
strongly sensitive to the regularity of the peaks, rather than just their 
strengths. In other words, by applying an autocorrelation we use the 
information that peaks are expected to be regularly spaced, whereas this information is 
disregarded when smoothing the power. The single strong peak close to 6\,mHz in the top panel of Figure \ref{fig01}, 
for example, is an artefact in the VIRGO photometry and produces a 
much more significant response in the smoothed spectrum than in the collapsed ACF.
Figure \ref{fig02} illustrates this further, using a power 
spectrum of a simulated Kepler star with low signal-to-noise. Compared to the smoothed power spectrum, 
the collapsed ACF shows a strong peak at the correct location of 
$\nu_{\rm max}$ and is clearly less sensitive to areas with white noise. 

The shape of the power envelope in other stars can be very 
different than the Sun \citep[e.g. for Procyon, see][]{arentoft} and hence might not be suitably modelled with 
a simple single Gaussian function. The collapsed ACF is less influenced by power 
asymmetries than a smoothed spectrum (see Figure 1, bottom panel), and an extension of the pipeline to 
include multiple Gaussians or different functions will be forthcoming. This will be of particular 
interest when analysing binaries in which both components show detectable power excess in the 
spectrum.

\figureDSSN{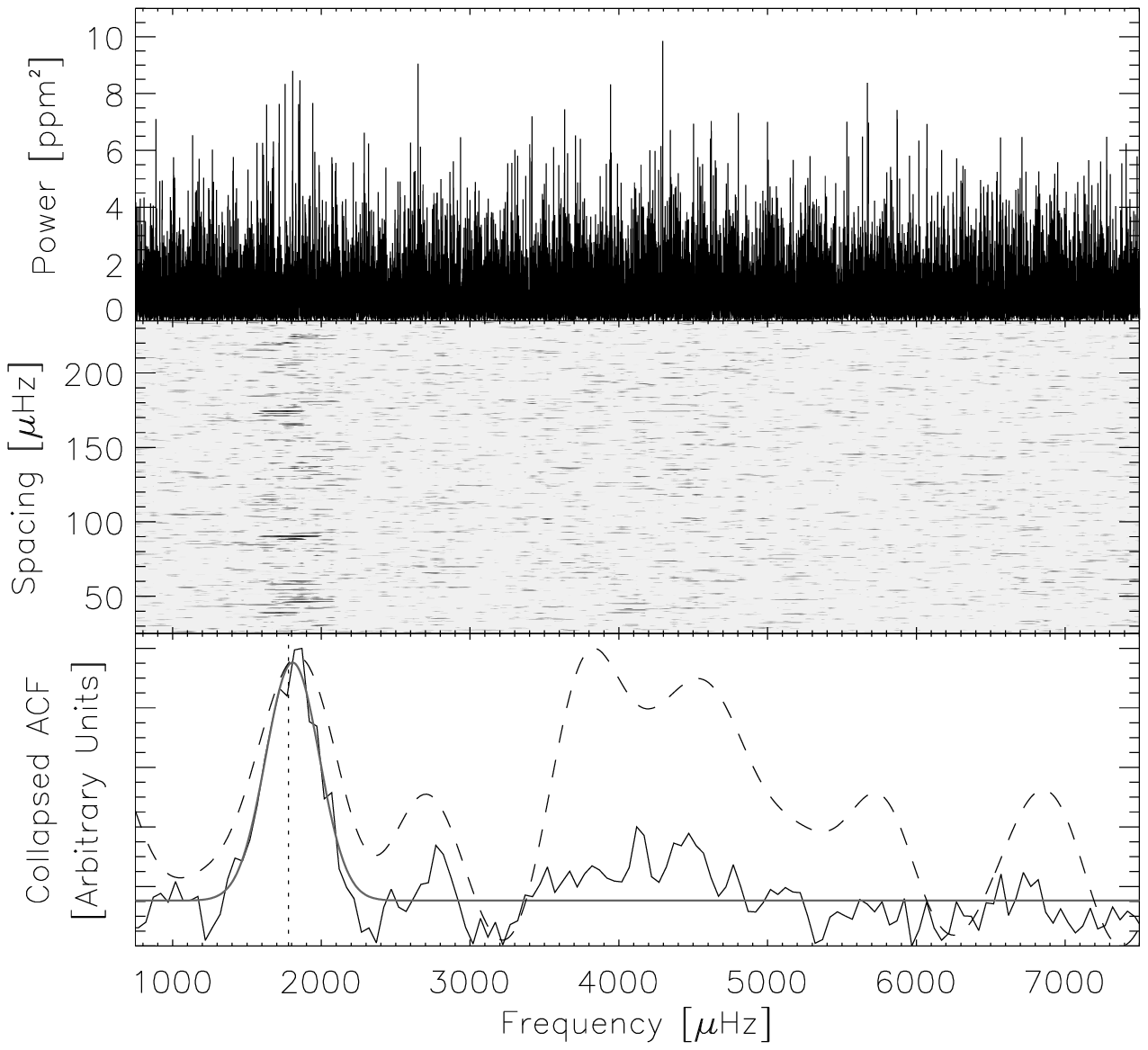}{Same as Figure \ref{fig01} but for a 1-month time series 
of a simulated Kepler star with low S/N. The vertical dotted line in the bottom panel is the true 
$\nu_{\rm max}$ value for this simulation. The large spacing was 88\,$\mu$Hz.}{fig02}{t!}{clip,angle=0,width=100mm}

\subsection*{3.2 Background modelling}
Modelling of power due to stellar background is widely done using a sum of power laws 
initially proposed by \citet{harvey}, with a revision of the power law exponent by 
\citet{aigrain}. Here, we use a mixture of the two versions that was originally 
suggested by \citet{karoff_phd} and has the form

\begin{equation}
P(\nu) = P_{n} + \sum_{i=0}^k \frac{4\sigma_{i}^{2}\tau_{i}}{1+(2\pi\nu\tau_{i})^{2}+(2\pi\nu\tau_{i})^{4}} \: ,
\label{equ:harv}
\end{equation}

\noindent
where $P_{n}$ is the white noise component, $k$ is the number of power laws used and $\sigma$ and $\tau$ 
are the rms intensity and timescale of granulation, 
respectively. The motivation behind this extended model is a more physically realistic 
interpretation of the stellar background. Instead of assuming a constant slope for the entire 
frequency range, it allows a shallower slope at low frequencies corresponding to turbulence (stellar activity)
and steeper slopes at higher frequencies corresponding to granulation \citep{nordlund}.

We determine $\sigma$ and $\tau$ using a least-squares fit to the power density spectrum. 
From a statistical point of view such an approach is questionable, 
since a raw power spectrum is not described by Gaussian statistics. We overcome this problem by 
smoothing the power spectrum using independent averages only \citep{garcia}, which allows a determination of 
parameter uncertainties. Alternative methods such as a
Bayesian approach using Markov-Chain Monte-Carlo simulations (T. Kallinger, private communication) 
could be used for more detailed studies of stellar granulation.

Regardless of the method of fitting, a pipeline relies on good initial conditions 
for a successful fit. To estimate such values, it is important to understand the rms intensity 
and, especially, the timescale of stellar granulation as a function of physical 
parameters. Based on numerical simulations of stellar surface convection, \citet{freytag} initially suggested 
that the linear size of a granule $l$ is proportional to the pressure scale height on the stellar surface:

\begin{equation}
H_{p}^{\rm surf} = \frac{l}{\alpha} \: .
\end{equation}

\noindent
Here, $\alpha$ denotes the mixing length parameter. Assuming that the cells move
proportional to the speed of sound $c_{s}$ \citep{svensson}, Kjeldsen \& Bedding (in 
preparation) show that, under the further assumption of adiabacity and an ideal gas, the granulation timescale 
can be expressed as

\begin{equation}
\tau_{\rm gran} \propto \frac{H_{p}^{\rm surf}}{c_{s}} \propto \frac{L}{T_{\rm eff}^{3.5} M} \: .
\end{equation}

\noindent
According to \citet{kjeldsen_amps}, this is inversely proportional to $\nu_{\rm max}$ and hence

\begin{equation}
\tau_{\rm gran} = \tau_{\rm gran,\sun} \frac{\nu_{\rm max,\sun}}{\nu_{\rm max}} \: .
\label{equ:tausc}
\end{equation}

This suggests that the timescale of granulation scales with the timescale of oscillations, which 
is plausible since both processes are tied to convection. Knowing $\nu_{\rm max}$ from the 
estimation performed in the previous section, granulation timescales can be scaled from the Sun without 
prior knowledge of stellar parameters.

\figureDSSN{fig03.eps}{Power density spectra of a 9-year VIRGO time series (top panel) and a 
60-day time series of HD49933 as observed by CoRoT (bottom panel). Dotted lines are the fitted 
individual power law components which, together with a 
white noise component (not shown), result in the final background model (thick solid lines). Dashed lines 
in the bottom panel show the initial guesses for the background fit using 
Equation \ref{equ:tausc}. Note the differences in y-axis scaling for each panel.}
{fig03}{}{clip,angle=0,width=100mm}

We tested this scaling method on HD\,49933, a star for which granulation 
and solar-like oscillations have been detected by CoRoT \citep{app_hd49933}. Figure 
\ref{fig03} compares the power density spectrum of HD\,49933 with the Sun, together with the 
individual power law components calculated using Equation \ref{equ:harv}. We assumed three 
components of stellar background in each spectrum: stellar activity at very low 
frequencies and two components due to different types of granulation. While the 
initial guesses for HD\,49933 (dashed lines) using Equation \ref{equ:tausc} yield a satisfactory 
final fit (dotted and solid lines), it is evident that the background in this star is somewhat different from 
the scaled Sun. This result confirms that granulation signatures in hotter stars are 
quite different from the Sun, as found by \citet{guenther} for a convection model of 
Procyon. We refer to \citet{ludwig_hd49933} for a detailed discussion of the granulation 
signal in HD\,49933 in the context of hydrodynamical simulations. 

Despite these differences for HD\,49933, Equation (\ref{equ:tausc}) appears to be a 
satisfactory approximation to provide initial values for background fitting and hence we implemented it in 
the pipeline. After the background has been successfully fitted, the power spectrum is 
corrected by dividing through the background model.

\subsection*{3.3 Estimation of $\Delta\nu$}
It is well known that the stochastic excitation and damping of solar-like oscillations causes 
series of peaks centred around the true frequency values in the power spectrum. To obtain a 
robust estimate of the average large frequency spacing, it is often helpful to divide the time series into subsets and 
co-add the corresponding power spectra, which leads to an average power spectrum with lower 
frequency resolution. In our pipeline, the background-corrected power spectrum is 
inverse-Fourier-transformed into the time-domain. The time series is then divided up into 
overlapping subsets (typically of 5 day length with a step size of 1 day) and power spectra 
of the individual subsets are co-added. This procedure forms a smoothed power spectrum. 
Figure \ref{fig04} compares the
original power spectrum with the background-corrected co-added power spectrum for a
simulated Kepler star.

As a next step, we repeat the power excess determination described in Section 3.1 using 
the background corrected co-added power spectrum. Using this final value for $\nu_{\rm max}$, we 
estimate the expected spacing by using the tight correlation between $\nu_{\rm max}$ and $\Delta\nu$ 
discussed by \citet{stello2009}

\begin{equation}
\Delta\nu_{\rm exp} \propto \nu_{\rm max}^{0.8} \: .
\label{equ:scale}
\end{equation}

Next, the autocorrelation of the power spectrum for the region $\nu_{\rm max} \pm 10\Delta\nu_{\rm exp}$ 
is calculated. Note that this width broadly agrees with the observed power excess in 
the Sun, and that we are at this stage only interested in deriving an average large frequency 
spacing over a large number of modes. Finally, we flag the five highest peaks in the autocorrelation, 
and fit a Gaussian function to the peak among the five which is closest to $\Delta\nu_{\rm exp}$, 
yielding the final determination of $\Delta\nu$. Figure \ref{fig05} shows a $\Delta\nu$ 
measurement of a simulated Kepler star for which the correct spacing does not correspond to 
the highest peak in the autocorrelation.

\figureDSSN{fig04.eps}{Top panel: Original power spectrum of a simulated Kepler star with a time 
base of one month. Bottom panel: Background-corrected and co-added power 
spectrum of the same star.}
{fig04}{t!}{clip,angle=0,width=100mm}

\figureDSSN{fig05.eps}{ACF of the background-corrected co-added power spectrum of a simulated 
Kepler star. Black crosses mark the five highest peaks. The grey line is a fit to the peak 
among the five which is closest to $\Delta\nu_{\rm exp}$ (vertical dotted line).}
{fig05}{h!}{clip,angle=0,width=100mm}

\subsection*{3.4 Uncertainties in $\nu_{\rm max}$ and $\Delta\nu$}
\label{sec:sim}
A crucial part of an automated pipeline is the ability to interpret the quality (or credibility) 
of the returned values. However, since we determine $\nu_{\rm max}$ and $\Delta\nu$ using least-squares fits 
to autocorrelation functions, determining 
reliable uncertainties is not straight forward. As pointed out by \citet{chaplin2008}, the formal 
uncertainties of such fits are strongly underestimated since the datapoints to which 
functions are fitted are highly correlated, and no proper weights (or uncertainties) can be 
assigned to individual datapoints. Additionally, the stochastic nature of solar-like oscillations 
introduces an intrinsic scatter of our measured $\nu_{\rm max}$ and $\Delta\nu$ values.

To overcome this, we followed the approach of \citet{chaplin2007} and performed simulations by 
producing synthetic time series. The inputs 
for each simulation were solar frequencies covering roughly twelve orders of $\ell=0-2$ 
modes taken from BiSON observations \citep{broomhall}. We modelled the amplitudes using a solar 
envelope derived from smoothing a power spectrum calculated from a 30-day subset of VIRGO photometry.
For simplicity, we assumed that all modes are intrinsically equally strong, but accounted for 
different spatial responses of $\ell=0-2$ modes according to \citet{kjeldsen2008}. We simulated 
the stochastic excitation and damping using the method of \citet{chaplin1997}, with a frequency 
independent mode 
lifetime of three days (i.e. solar). Each time series consisted of the same sampling and time base as 
the simulated asteroFLAG stars, and white noise was added to each synthetic time series. 

We performed simulations with different input amplitudes and frequencies to resemble a range of 
stellar evolutionary states, and with different S/N corresponding to a variety of stellar magnitudes. We made 
100 realizations, including stochastic excitation and white noise for each set of input parameters. 
%The solar BiSON frequencies were shifted by -3098\,$\mu$Hz (the BiSON frequency of maximum amplitude) and 
%then, for a given $\nu_{\rm max}$, multiplied by $\Delta\nu_{\rm exp}/\Delta\nu_{\sun}$, with 
%$\Delta\nu_{\rm exp}$ given by Equation \ref{equ:scale}. Finally, all frequencies were shifted to 
%the input value $\nu_{\rm max}$. 
The resulting light curves were then 
analysed by the pipeline, and the standard deviations of the determined 
values for $\nu_{\rm max}$ and $\Delta\nu$ were taken as the true uncertainties. 

\figureDSSN{fig06.eps}{Top panels: Scatter of $\nu_{\rm max}$ (left) and $\Delta\nu$ (right) 
from simulations as a function of S/N. Darker colours correspond to higher input values of $\nu_{\rm max}$ 
and $\Delta\nu$, respectively. 
Bottom panels: Correction factors for formal 
uncertainties as a function of S/N and input value.}
{fig06}{tbp}{clip,angle=0,width=100mm}

The results of the simulations are shown in Figure \ref{fig06}. 
The top panels show the scatter in $\nu_{\rm max}$ and $\Delta\nu$ as a function of S/N. We see 
that the scatter of $\nu_{\rm max}$ is greater for higher values of $\nu_{\rm max}$ (darker symbols), 
while the scatter in $\Delta\nu$ is almost independent of $\Delta\nu$. This is expected, since 
the power excess hump used to determine $\nu_{\rm max}$ becomes broader for higher input values of 
$\nu_{\rm max}$ and hence the absolute deviation increases. On the other hand, the peak in the 
autocorrelation used to determine $\Delta\nu$ will remain about the same because it is 
determined by the frequency resolution and mode lifetime, which are the same for all simulations. 
We note that the scatter reaches a constant level 
for high S/N, and the maximum precision with which $\nu_{\rm max}$ and $\Delta\nu$ can be determined 
are $\sim$10\,$\mu$Hz and $\sim$0.1\,$\mu$Hz, respectively. The latter value is in good agreement 
with the uncertainties reported in the first asteroFLAG exercise \citep{chaplin2008}.

The ratio between 
these values and the formal uncertainties as determined by the least-squares fit give a look up table of 
correction factors which we use to convert the formal uncertainties into more realistic values.
To obtain smoothly varying correction factors, we fitted power laws to the results of the 
simulations. These are shown in the bottom panels of Figure \ref{fig06}. As expected, the factors increase 
for higher S/N, i.e. the least squares fit underestimates uncertainties more for higher signal because 
the correlation between fitted points is higher. Towards the detection limit at low S/N, this 
trend quickly reverses and formal uncertainties must be scaled with high factors to 
accommodate the large uncertainty due to high noise levels. We note 
that at S/N values around 10 and lower, the $\Delta\nu$ uncertainties determined by the least-squares fit are 
in fact overestimated, with correction factors $<$\,1. Considering that our simulations are 
simplified compared to real data and therefore the scatter at low S/N values is likely 
underestimated, we disregard this effect and do not downscale formal uncertainties.

We note that the uncertainty correction presented here will also be applicable to real Kepler stars, 
with slight adaptations for different sampling and observing lengths. In preparation for this, 
we test our uncertainties using simulated Kepler stars, which will be presented in the next section.

\section*{4 Application to simulated Kepler observations}

We applied the pipeline, as described in the previous section, to 1936 simulated Kepler stars 
discussed in Section 2. 
To verify the values returned by the pipeline, we calculated ``true'' values of 
$\nu_{\rm max}$ and $\Delta\nu$ as follows: Using the stellar mass, 
luminosity and effective temperature of the input model, we calculated $\nu_{\rm max,true}$ 
using the scaling relation by \citet{kjeldsen_amps}. To determine $\Delta\nu_{\rm true}$, we 
first determine the input model frequency closest to $\nu_{\rm max,true}$. We then fitted a linear 
regression to ten orders of the same degree around the frequency of maximum power, and used the slope 
to estimate the frequency spacing \citep{nearsurf}. This was done separately for modes of $\ell=0-2$, and the 
final value of $\Delta\nu_{\rm true}$ is a weighted mean of the three spacings, 
with weights corresponding to the spatial responses as given by \citet{kjeldsen2008}. Note that 
for more evolved stars ($\Delta\nu < 70\,\mu$Hz), no reliable model frequencies were 
available and hence the scaling relation by \citet{kjeldsen_amps} was used to calculate $\Delta\nu_{\rm true}$.

To investigate systematic effects in our uncertainty simulations from section 3.4, we 
repeated these simulations for noise-free realizations and compared $\nu_{\rm max}$ and $\Delta\nu$ 
returned by the pipeline to $\nu_{\rm max,true}$ and $\Delta\nu_{\rm true}$ for these simulations. We 
found that on average the determined $\nu_{\rm max}$ values are $\sim$\,1\% higher and that the determined 
$\Delta\nu$ values are $\sim$\,0.05\,\% lower than the input values. Both effects are 
easily understood: In our uncertainty simulations, as well as the Kepler simulations, a 
solar oscillation profile was assumed. While the amplitudes in this profile have positive asymmetry,  
the large spacings increase towards higher frequencies. The collapsed ACF used to determine $\nu_{\rm max}$ 
is sensitive to regular peak spacings, and hence overestimates $\nu_{\rm max}$ compared to our 
definition of $\nu_{\rm max,true}$, which is not equal to the center of the solar envelope. 
The single ACF used to measure $\Delta\nu$ is influenced by the peak power, and hence 
underestimates $\Delta\nu$ compared to our definition of $\Delta\nu_{\rm true}$, which is the 
mean spacing across the envelope independent of amplitude. For this application, we account for both effects by 
multiplying the measured $\nu_{\rm max}$ values by 0.99 and the measured $\Delta\nu$ values by 
1.0005.

We now proceed to the main results by comparing the measured values of the 1936 Kepler stars 
with the true values. Figures \ref{fig07} and \ref{fig08} display the differences between the true values (as 
defined above) to the quantities measured by the pipeline for $\nu_{\rm max}$ and 
$\Delta\nu$, respectively. In this comparison we have eliminated all stars with relative uncertainties 
greater than 10\%. Furthermore, we disregard extreme outliers by considering only measurements for which 
the absolute difference of $\Delta\nu$ and $\Delta\nu_{\rm exp}$ is lower than 0.1 $\Delta\nu_{\rm exp}$, 
and the absolute difference of $\nu_{\rm max}$ and $\nu_{\rm max,exp}$ is lower than 0.5 $\nu_{\rm max,exp}$. 
$\nu_{\rm max,exp}$ is calculated using the scaling relation by \citet{kjeldsen_amps} with the Kepler Input Catalog 
parameters for these stars. Of the entire sample, 52\% of the $\nu_{\rm max}$ and 48\% of the $\Delta\nu$ 
measurements fulfill these criteria.

\figureDSSN{fig07.eps}{Top panel: Differences between true and measured values for $\nu_{\rm max}$ 
for all measurements with an uncertainty precision lower than 10\%. The inset shows the comparisons 
on a larger scale to display measurements with 
larger deviations. Bottom panel: Mean estimated uncertainties of the measurements (solid line) compared to the 
actual scatter of measurements and true values (dashed line). In both panels, the dashed-dotted and dotted 
lines display uncertainty limits necessary to constrain stellar radii to 1\% and 2\%, 
respectively.}
{fig07}{tbp}{clip,angle=0,width=100mm}

\figureDSSN{fig08.eps}{Same as Figure \ref{fig07} but for $\Delta\nu$.}
{fig08}{tbp}{clip,angle=0,width=100mm}

The largely symmetric distributions for both $\nu_{\rm max}$ and $\Delta\nu$ in the top panels
of Figures \ref{fig07} and \ref{fig08} suggest that systematic effects caused by the methods of the pipeline have 
been mostly removed or corrected. We suspect that the slight negative trend for high values of 
$\Delta\nu$ is due to the fact that $\Delta\nu_{\rm true}$ is calculated based on a fixed number of model 
frequencies, which 
underestimates $\Delta\nu_{\rm true}$ compared to simulated Kepler stars where less low-frequency 
modes might be visible. 
As discussed in section 3.4, the scatter in $\nu_{\rm max}$ 
gets considerably larger as the 
power excess shifts to higher frequencies and the oscillation amplitudes become smaller. 
The bottom panels show the mean uncertainties and the rms of measured minus true values. In both cases, the 
curves are for the most part overlapping. The exception are low values of $\Delta\nu$ for which 
the uncertainties still seem considerably underestimated. We suspect that this is partially connected to 
the fact that for these stars $\Delta\nu_{\rm true}$ was calculated using a scaling relation 
rather than the actual model frequencies (see above).

An important application of asteroseismology within the Kepler mission will be to determine radii of 
of exoplanet host stars. As noted by \citet{chaplin2008}, 
a radius determination to a precision of 1\% requires a relative uncertainty of 0.15\,\% 
on the large frequency spacing. Formally, this corresponds to a 2\% relative uncertainty on 
$\nu_{\rm max}$. These precisions are indicated in Figures \ref{fig07} and \ref{fig08} by
dashed-dotted lines. We also show the precisions required for a more pessimistic 
radius precision of 2\% (dotted lines). The scatter of the measured values and the mean uncertainties shows 
that the 1\% limit should be achievable for a considerable number of stars 
up to $\Delta\nu\sim$\,100\,$\mu$Hz and $\nu_{\rm max}\sim$\,2000\,$\mu$Hz, and even higher values 
if the criterion is relaxed to a radius precision of 2\% or 3\%. These results therefore indicate an 
optimistic outlook for the automated analysis of Kepler asteroseismology stars, despite the fact that the 
time base of the initial survey will only be about 30 days.

\figureDSSN{fig09.eps}{H-R diagram of all 1936 stars in the test sample based on the model 
parameters. Black circles show stars for which the pipeline returned correct large frequency 
spacings precise enough to determine the stellar radius to a precision of 1\%.}
{fig09}{tbp}{clip,angle=0,width=100mm}

To analyse these results in terms of stellar evolution, Figure \ref{fig09} shows a H-R 
diagram based on the model parameters of all 1936 sample stars. The majority of the 
test sample was made up of main sequence and sub-giant stars. As expected, a large part of the 
best $\Delta\nu$ determinations (with uncertainties allowing a radius determination to 1\% precision) 
were made in stars with relatively high luminosities. The 
relative lack of detections for the evolved sub-giants compared to the rest of the sample points to a
problem of our pipeline with handling signal at the very low frequencies. Quite 
surprisingly, the results also indicate that precise $\Delta\nu$ determinations will be possible 
for a number of cool, low-mass main-sequence stars (which represent the datapoints at very high 
$\nu_{\rm max}$ and $\Delta\nu$ in Figures \ref{fig07} and \ref{fig08}).

\section*{5 Summary \& Conclusions}

We have described an automated analysis pipeline to extract global oscillation parameters for a 
large number of stars. We demonstrated that the use of a collapsed autocorrelation 
function is a sensitive tool to find the location of excess power. We further showed that a determination of 
$\nu_{\rm max}$ can be used to scale granulation timescales in order to model the background contribution 
in the power 
spectrum. To obtain robust uncertainty estimates on $\nu_{\rm max}$ 
and $\Delta\nu$, we have performed realistic simulations of solar-like oscillations as a 
function of S/N and $\nu_{\rm max}$, and derive correction factors which are necessary to convert 
least-squares uncertainties derived from correlated data to realistic uncertainties. Our 
simulations indicate that for a one-month time series with one-minute sampling, the maximum 
precision with which $\nu_{\rm max}$ and $\Delta\nu$ can be determined are $\sim$10\,$\mu$Hz and 
$\sim$0.1\,$\mu$Hz, respectively.

The automated pipeline was applied to a sample of 1936 simulated stars representing targets of the 
Kepler asteroseismic survey phase. We show that our scaled uncertainties are reliable for all values of 
$\nu_{\rm max}$, but seem to be significantly underestimated for $\Delta\nu < 50\mu$Hz. While we 
suspect that this is mostly due to the fact that $\Delta\nu_{\rm true}$ could not be calculated 
accurately from model frequencies for evolved stars, the results show that in general some 
modifications of the code are needed for processing stars that pulsate at 
low frequencies ($<$\,500\,$\mu$Hz). The further development of the pipeline, in particular with 
respect to the background modelling, will focus on this adaptation to process red giant stars.

The comparison of real and measured values showed that in 70\% and 60\% of all cases, $\nu_{\rm max}$ and 
$\Delta\nu$ were recovered within 10\,\% of the true value, respectively. Using 
the estimated uncertainties to eliminate measurements with too large uncertainties and disregarding 
extreme outliers, these numbers drop to roughly 50\%. The scatter of the measured 
values around the input values and the mean uncertainties agrees well for this sample, and 
indicate that for at least 20\% of the stars $\Delta\nu$ can be determined with a precision 
sufficiently high to infer stellar radii to 1\% accuracy. Plotting these stars in an HR 
diagram suggests that their distribution is quite diverse, including low-mass main sequence stars as 
well as evolved sub-giants.

\acknowledgments{This work benefited from the support of the International Space Science
Institute (ISSI), through a workshop programme award. It was also partly
supported by the European Helio- and Asteroseismology Network (HELAS), a
major international collaboration funded by the European Commission's
Sixth Framework Programme. WJC acknowledges the support of the UK
Science and Technology Facilities Council (STFC). DH likes to thank Rafael Garcia, 
Christoffer Karoff, Stephen Fletcher, Michael Gruberbauer and Thomas Kallinger for interesting and 
fruitful discussions.}

%\newpage

\bibliographystyle{aa}
\bibliography{refs}

%\References{
%\rfr Author 1, \& Author 2 Year, Journal-Abbrev., Issue-No, first page of article
%\rfr Author 1, Author 2, \& Author 3 Year, Journal-Abbrev., Issue-No, first page 
%\rfr Author 1, Author 2,  Author 3, \& Author 4 Year, Journal-Abbrev., Issue-No, first page
%\rfr(for more than 4 authors:) 
%\\ Author 1, Author 2, Author 3, et al. Year, Journal-Abbrev., Issue-No, first page
%}

\end{document}

%% file: CoAst_regularlayo
\renewcommand{\chaptermark}[1]{\markboth{#1}{}}

\newcommand{\kopf}{\small\itshape Comm. in Asteroseismology \\ Vol. number, publication date (will be inserted in the production process)}

\newcommand{\Authors}[1]{\begin{center}\normalsize\bf\sf #1 \end{center}}

\renewcommand{\author}[1]{\begin{center}\normalsize\bf\sf #1 \end{center}}
\newcommand{\Address}[1]{\begin{center}\small\sf #1 \end{center}}

\renewenvironment{abstract}{\section*{Abstract}\normalsize\sf}{}

\newcommand{\chapterCoAst}[3]{\chapter[\sf\normalsize #1\\ \footnotesize \hspace*{5mm}by #2 \sf\normalsize][]{#1\\}\rhead[\fancyplain{}{\sf\footnotesize \center{#3}}]{\fancyplain{}{\sffamily\thepage}}\lhead[\fancyplain{\kopf}{\sffamily\thepage}]{\fancyplain{\kopf}{\sf\footnotesize \center{#2}}}}

\newcommand{\figureDSSN}[5]{\begin{figure}[#4]
\centering
\includegraphics*[#5]{#1}
\caption{#2}
\label{#3}
\end{figure}}

\renewcommand{\chaptername}{}

\newcommand{\acknowledgments}[1]{\vspace*{5mm}\noindent  \textbf{Acknowledgments.} #1}